\documentclass{ws-ijmpc}
\usepackage[utf8]{inputenc}
\usepackage{algorithmic}

\begin{document}

\markboth{Michał Żabicki}
{OpenCL/OpenGL approach for studying active Brownian motion}

\catchline{}{}{}{}{}

\title{OpenCL/OpenGL approach for studying active Brownian motion}

\author{Michał Żabicki}

\address{M. Smoluchowski Institute of Physics, Jagiellonian University, Reymonta 4, 30-059 Krakow, Poland\\
michal.zabicki@uj.edu.pl}

\maketitle

\begin{history}
\received{Day Month Year}
\revised{Day Month Year}
\end{history}

\begin{abstract}
This work presents a methodology for studying active Brownian dynamics on ratchet potentials using interoperating OpenCL and OpenGL frameworks.

Programing details along with optimization issues are discussed, followed by a comparison of performance on different devices.

Time of visualization using OpenGL sharing buffer with OpenCL has been tested against another technique which, while using OpenGL, does not share memory buffer with OpenCL.
 
Both methods have been compared with visualizing data to an external software - gnuplot.
 OpenCL/OpenGL interoperating method has been found the most appropriate to visualize any large set of data for which calculation itself is not very long.

\keywords{active Brownian motion; stochastic differential equations; OpenCL --- OpenGL interoperation;  GPGPU}
\end{abstract}

\ccode{PACS Nos.: 87.10.Mn, 07.05.Tp, 89.20.Ff}

\section{Introduction}
\label{intro}

Over the last few years General Programming on Graphical Processing Units (GPGPU) has started to spread in all the areas where time saving and performance of calculations is crucial.
Scientific simulation is the perfect example \cite{Rossinelli:2010p101,Stone:2010p167,Sundholm:2010p184}.

Motion of molecular motors has been simulated using the different approaches and models \cite{Zhang:2009p1072,Derenyi:1996p860,Parker:2009p1446,Bowling:2009p1244,Bier:2011p502}.
The inherent feature of modeling stochastic processes is the noise, i.e. non-systematic fluctuating increments of the process.
To obtain the most probable behavior of the system it is necessary to repeatedly perform calculations and then calculate their mean values based on stochastic properties of the ensemble.
On the other hand, chaotic behavior of models with respect to initial parameters requires simulation for different sets of both initial conditions and model parameters.
All of those tasks can be easily divided into separate threads --- problem is highly parallel.

Most of high performance computing (HPC) in science is performed on clusters --- sets of single or multiple-core processing units connected in a network.
The workflow consists of the preparation of a source code, its transmission over the network using a cluster controller, putting a {\it job} in a {\it queue}, some waiting time, the execution of the parallel program and finally the acquisition of results over the network.

Fast as this approach may seem, it has several drawbacks.
Firstly, for many simulations the benefit of faster calculations, if one includes time for "sending", "waiting" and "acquisition", might be rather insignificant.
 Secondly, the cost of every computer forming a cluster is much higher than the cost of even high-end graphical processing unit (GPU).
Every GPU itself is a multiprocessing unit - even a low-end Nvidia GeForce 320M GPU found on 11'' Apple MacBook Air laptop has 48 cores, while SU9400 Intel processor (CPU) found on the same device offers only 2 cores \cite{mbaspecs}.

The most popular technique for performing GPGPU nowadays is NVidia's proprietary technology CUDA.
While very efficient, it is vendor agnostic --- it runs only on modern Nvidia graphical cards.
In 2008, Khronos Group consortium created first specification of OpenCL which has been declared an "open standard for programming heterogeneous data and task parallel computing across GPUs and CPUs" \cite{Munshi:2011p489}.

Heterogeneous approach allows researchers to write programs managed by {\it hosts} that can be run on different OpenCL {\it devices}, even the ones where GPUs cannot be used for OpenCL calculations.
On that occasion, simulations are performed on all of the CPU cores.
Performance is worse, yet calculation can be done without a single change of the source code.
OpenCL's parallel nature also allows using all the CPU processing power without any special multithreading programming --- tasks are dispatched by OpenCL into different threads to all available processor cores.

One of the downsides of GPGPU is that both writing data to GPU and reading data back to CPU  is carried over a relatively slow system bus (on PCI-E2 x8 bus Host (CPU) --- Device (GPU): 2.4-2.5 GB/s and Device (GPU) --- Host (CPU): 1.9-2.0 GB/s \cite{Anonymous:up4-Mhma}).
It takes the same time to transfer data from CPU to GPU  and to carry more than a few operations on the GPU on the same set of data.
This behavior will be evaluated later in the text.
That said, what is crucial for good overall simulations performance is to carry GPU - CPU data transfer only when it is really necessary.

At this point, there is OpenGL, a technology developed over 20 years ago by Silicon Graphics and now maintained by the Khronos Group, which as already mentioned, is responsible for developing OpenCL.

OpenGL is a standard specification for writing programs that produce computer graphics.
Creating an image on a monitor consists of setting environment, sending instructions to the GPU to execute them and finally show the result on the screen.
For example,  rotation of an object is performed by sending a simple command to rotate by some angle and over some axis, while the calculations of all coordinates are made by GPU, without CPU burden.

OpenCL even in its initial specification mentions the possibility of integration with OpenGL.
Sending the results, which are already in the GPU memory, to the CPU and re-sending them back to the GPU to visualize them seems an obvious waste of both time and device processing resources.

In this paper, I present an approach which integrates simulation and presentation of the results on the same processing unit, GPU.
If for some reason OpenCL cannot be run on GPU, it still can work on CPU --- much more efficiently than by simple execution of a single-thread simulation.
Because of the little performance impact, results can be presented after trajectories for a given sets of parameters are calculated.
Over time, results are averaged and a researcher can terminate computation when he finds results sufficient to validate or invalidate his thesis.

Specific problems of active Brownian motion (ABM) simulations using OpenCL are explained in a subsequent section.
Comparison of performance on different hardware and software systems is also provided.

\section{A physical problem and programming issues}

The sample ABM simulation is a model of Langevin dynamics coupled to energy depot \cite{Fiasconaro:2008p325}.
This and following models \cite{Zabicki:2010p59,Zabicki:2010p264} show a rich dynamical behavior of transfer problems on ratchets.
The efficiency of models depends strongly on both initial and running parameters.
Studying their behavior requires a number of simulation runs for different sets of conditions.
Stochastic nature calls for repeated simulations of trajectories which correspond to realizations of the same dynamic process.

In the aforementioned model (ABM) the depot energy $e(t)$ changes in time $t$ according to equation: 
The mechanical energy of motion $v$ comes from the energy flow from the depot which depends on the coupling parameter $d$: 
 \begin{equation}
\dot{e} = q - c e(t) - d e(t) v^2(t),
 \label{ereservoir}
 \end{equation}
 where $c$ is a dissipation rate of energy in the container $q$ and the mechanical energy of motion $v$ comes from the energy flow from the depot which depends on the coupling parameter $d$.

Equation of motion incorporates $\gamma$ as a velocity-dependent friction, external force $F_0$ and white Gaussian noise of intensity $\sqrt{2D}$:
  \begin{eqnarray}
 \dot{v}=F_0+de(t)v(t)-\gamma v(t)-U'(x)+\sqrt{2D}\xi(t) {v}
 \label{velocity}
 \end{eqnarray}
where $U(x)$ is a ratchet potential of the height $h$:
\begin{eqnarray}
\nonumber U(x)=h[0.499-0.453(sin(2\pi(x+0.1903))\\ +\frac{1}{4}(sin(2\pi(x+0.1903)))].
\label{pot}
\end{eqnarray}

One of the main topics studied recently (see Refs. \cite{Zabicki:2010p59,Zabicki:2010p264}), has been the problem of efficiency related to opposing external force $F_0$.
One of the methods that illustrates the problem are plots of average velocity as a function of the force $F_0$.
This relation is used in this work to compare the performance for various software and hardware setups.

Previous approaches utilized to check similar relations were based on serial multiple program executions, with parameters governed by Perl script.
In the OpenCL method, several instances of the same program, different only by force parameter $F_0$ are run simultaneously.
Every instance of that program is called a {\it work item} in OpenCL.
Over time, results for given parameters are averaged to maintain reliable result, independent of a given set of random numbers.

For every set of parameters OpenCL {\it kernel}, as in algorithm (see Fig. \ref{alg1}), is executed.
\begin{figure}[htbp]
\caption{Pseudo kernel}
 \label{alg1}
\begin{algorithmic}
\STATE read input parameters from global memory
\STATE fill local memory with parameters
\FOR{$i=0$ \TO $T$} 
\STATE do Marsaglia xorshift
\STATE do Box-Muller transformation
 \STATE calculate new velocity and energy (as in Eqs. \ref{ereservoir},\ref{velocity},\ref{pot})
 \ENDFOR 
\STATE assign local parameters to global memory
 \end{algorithmic}
 \end{figure}
 
After the kernel finishes its calculations, different techniques are used to visualize the results.
Their performance and the differences between them are discussed in Sec. \ref{interop_section}.

Solving stochastic differential equations requires generating noise (in Eq. \label{velocity} $\xi(t)$ is understood as a source of a Gaussian white noise), whose computer equivalent is a set of pseudo-random numbers.

The programmer can either fill OpenCL buffers with random numbers provided by host RNG or write OpenCL implementation of existent RNG algorithms.
The downside of the first approach is the time it takes to pass numbers to GPU --- in case of OpenCL/OpenGL it is almost always more efficient to send an initial number and a set of routines which would be executed on the GPU.
In this work, I have chosen the second way of generating random numbers, as in \cite{Januszewski:2010p94}.
Marsaglia xor-shift algorithm \cite{Marsaglia:2003p197} has been used and because white Gaussian noise needs normal number distribution instead of unitary, Box-Muller transformation has also been applied.

The other problem that arises when one writes a program running on the GPU is the precision issue.
Using double precision (DP) variable types (like {\it double}) is a standard for most scientific calculations.
While the newest graphic cards available on the market allow DP usage, for the sake of compatibility with the ones that do not have this possibility, kernels should use single precision (SP) variable types like {\it float}.
SP programs can be as precise as the DP ones as long as the technology drawbacks are overcome in a correct manner \cite{Langou:2007p406}.
This includes avoiding adding very small numbers to very big ones.

While offering less flexibility in setting high precision data types, OpenCL specification encourages programmers to use vector data types.
On certain graphic cards {\it float4} (consisting of four float numbers) is the "natural" data type and keeping the same structure of data may increase performance.
It should be also noted that, if executed on modern CPUs, using vector data types can also be beneficial in shortening calculation time \cite{Rosenberg:2010p490}.

\begin{figure}[htbp]
\begin{center}
\includegraphics[]{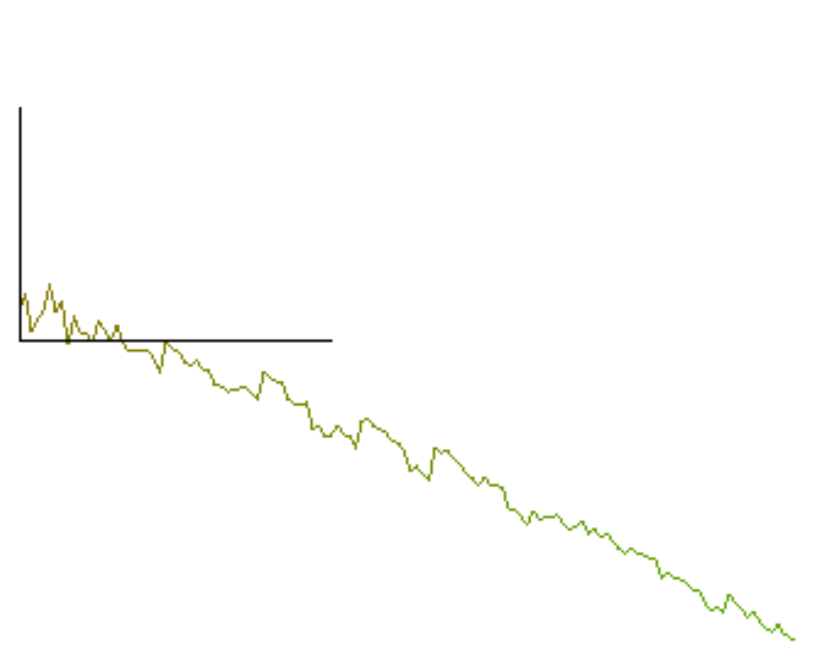}
\caption{Example plot of tested program.
Mean velocity $\langle v \rangle$ is plotted against opposing force $F_0$ (see details in the text).}
\label{example_plot}
\end{center}
\end{figure}

\section{Performance issues}

\begin{figure}[htbp]
\begin{center}
\includegraphics[scale=0.32]{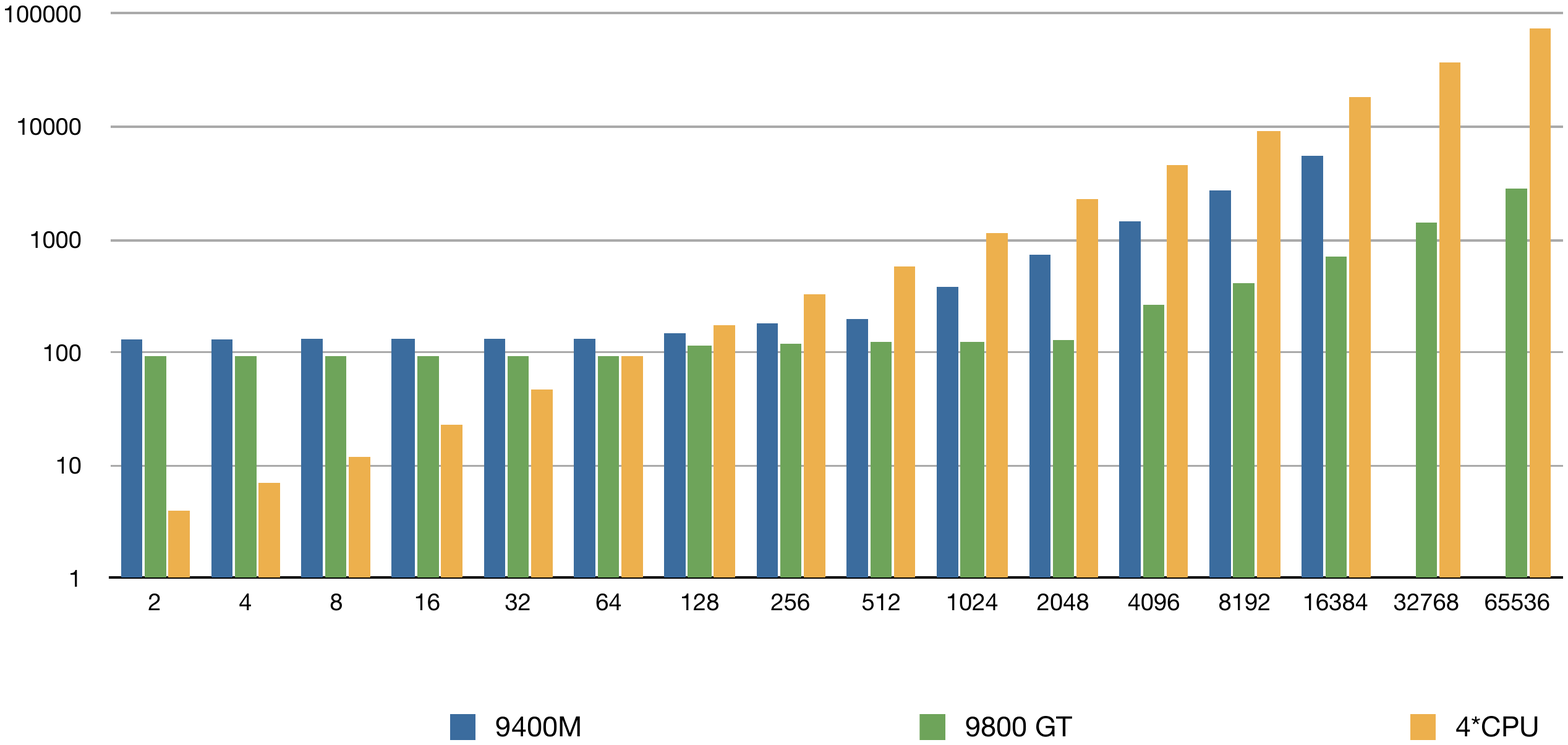}
\caption{Comparison of calculation time, depending on used hardware and number of simulation points.
{\bf Lower = better} }
\label{fig_of_results}
\end{center}
\end{figure}

A test program has been compiled and run on various Mac OS X 10.6 capable computers.
Throughout the test run all the other user visible applications have been shut down.
Every test sequence consisted of running the program for 20 steps for every power of 2 from $2^0$ to $2^{16}$ starting parameters, i.e. work items.
Then, the last fifteen steps have been taken into account and the average has been plotted (Fig. \ref{fig_of_results}).

Depending on a chosen hardware, time used for calculations varies substantially.

While having multiple cores (from few up to thousands), graphic cards suffer delays in every situation which requires transferring data to and from it to the CPU.
GPUs usually work on lower frequency clocks.
On the other hand, easy memory access and high frequency clocks would not overcome the main CPU drawback - low number of processing cores (from one to six in most cases).

\subsection{Performance on various simulation setups}

Preliminary tests of my program show that, depending on the number of parallel tasks to compute, GPU can be slower or faster, comparing to CPU.
In the case of a traditional central processor calculation, on 4-core Q6600 it always takes twice the time to compute twice larger set of simulation points.

On the other hand, it does not take significantly longer for the GPU to compute more simulation points until certain threshold number is reached.
The latter is a generic characteristic of a given graphic card.
Most obvious threshold should be GPU's core count.
As the test results show, it is not always true.

Nvidia GeForce 9400M is a popular graphic card used in laptops until last year.
It has 16 processing cores and uses up to 512 MB of system memory (in case of this test - DDR3).
While one can expect no difference in time of calculations until number of simulation point reaches number of cores and increase from thereon, actually this threshold occurs at the number of 128, which is 8 times the core count.

Introduced in 2008, Nvidia GeForce 9800 GT is a PCI-Express standalone card.
In this test, version with 512 MB GDDR3 internal memory has been used.
It has 112 cores working at 1500 MHz processor clock.
Here, there is a clear threshold corresponding to the number of cores when the time of a single simulation starts to increase.
However, unless the number of simulation points exceeds 2048 every doubling of required simulation points (say, an increase from $2^8$ to $2^9$) does not result in doubling of the run time.

In other words, it means that there is no difference in simulation time whether or not researcher calculates result for one or a thousand parameters --- as long as the upper limit is under the threshold, which is dependent on the graphic card used.

\begin{figure}[htbp]
\begin{center}
\includegraphics[scale=0.30]{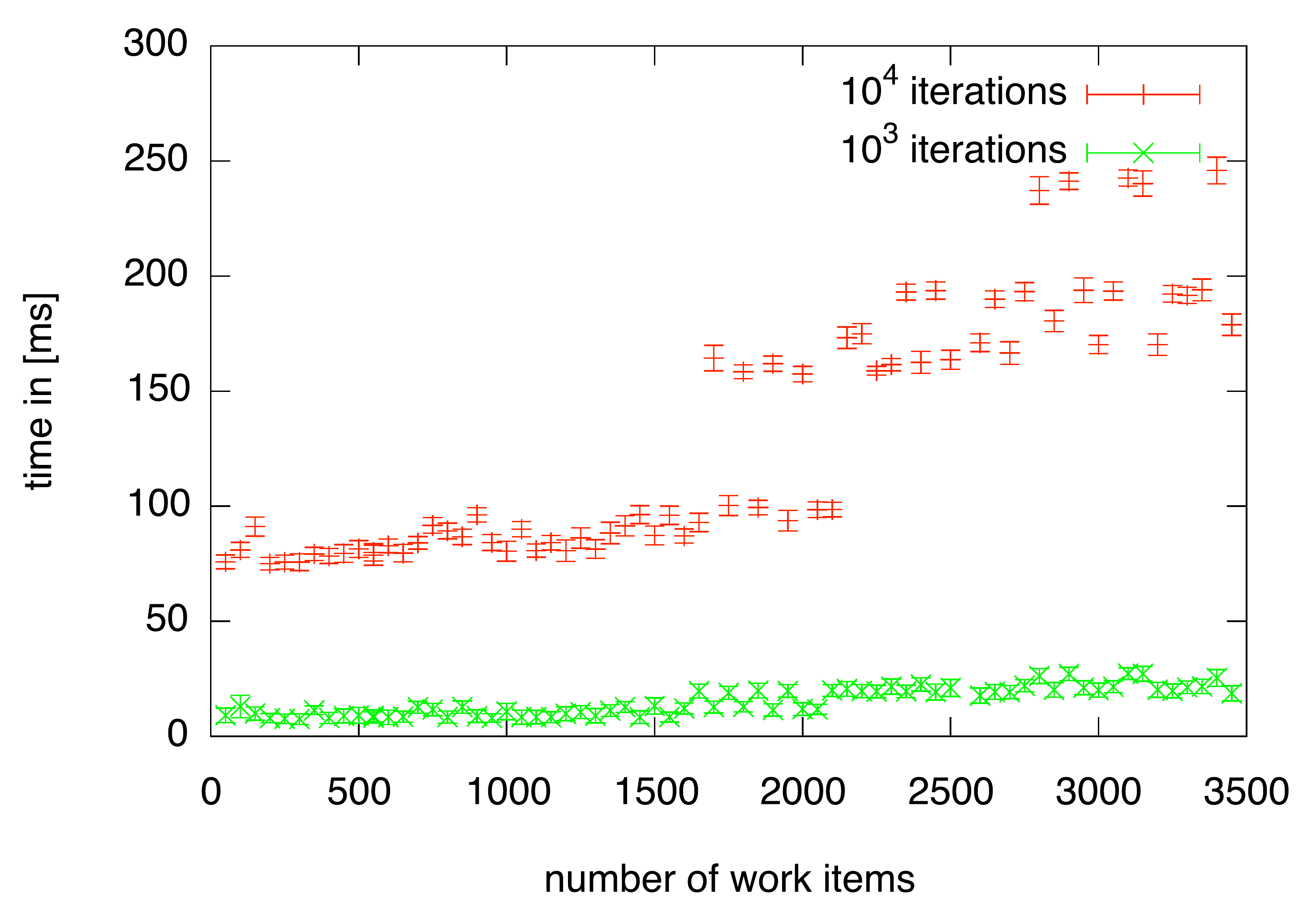}
\caption{Time of OpenCL calculations carried on GeForce 9800 GT GPU against number of work items.
Band-like structure could be notice instead of linear rise of calculation time.}
\label{9800gt}
\end{center}
\end{figure}

To further study OpenCL GPU performance, I have carried out more detailed calculations.
Between 50 and 3500 work items, with the increment of 50, I have measured the time of a one calculation step.
Because for every work item number calculations have been performed at least 20 times, he evaluation of the average and standard deviation has been carried out.
The procedure has been done for $10^4$ and $10^3$ iterations steps and for zero steps.
The latter helped to measure the offset time for initial variables' transfer from global to local memory as well as final transfers from local to global memory and visualization (performance of which is discussed in the second part of the article).
Subtracting offset time from "regular" calculations' time provided a more detailed view of the matter --- result can be found in the Fig. \ref{9800gt}.

The most prominent feature that can be observed in Fig. \ref{9800gt} is the band structure.
There is almost no rise in time of calculation until number of work items slightly passes 2000.
This is similar observation like in the previous plot.
 
What distinguishes it, is the situation repeats for what is happening over 2000 work items.
It can be said that GPU operates in certain regimes of performance, and crossing the thresholds results in non-linear rise of calculation times, i.e. while on CPU time of calculation is a linear function of work items (work to be done), on the GPU time of calculation rather can be explained as floor- or ceiling-like functions.

Closer look at the plot in the Fig. \ref{9800gt} can reveal that regardless of the number of iterations, bands occur in similar places --- for the same number of work items.
In the authors opinion, for a given graphic card there exists a maximum number of work items that can be done at the same time without any significant performance impact.
9800 GT GPU consists of 14 cores for each 8 streaming processors are provided and that makes 112 processors to operate at the same time.
Every core operates in 32 {\it warps} that help hide latencies of the memory.
It appears that this about-2000 is the threshold after reaching which GPU has to employ extra cycle to utilize all the work items.
That situation seems to reappear for the aforementioned threshold multiples.

For some near-the-threshold regions one can see that the time of calculation sometimes lies on the longer time band.
One of the explanations could be that GPU has been used at the moment of simulation for some other, most probably, system task.

\subsection{OpenCL/OpenGL interoperation performance}
\label{interop_section}

In this part, I will analyze the impact on performance, when intermediate simulation steps are shown to the user.
In all tests the time for one cycle (gathering initial parameters from memory, actual calculations and on-screen presentation of results) has been counted in microseconds.

\begin{figure}[htbp]
\begin{center}
\includegraphics[scale=0.15]{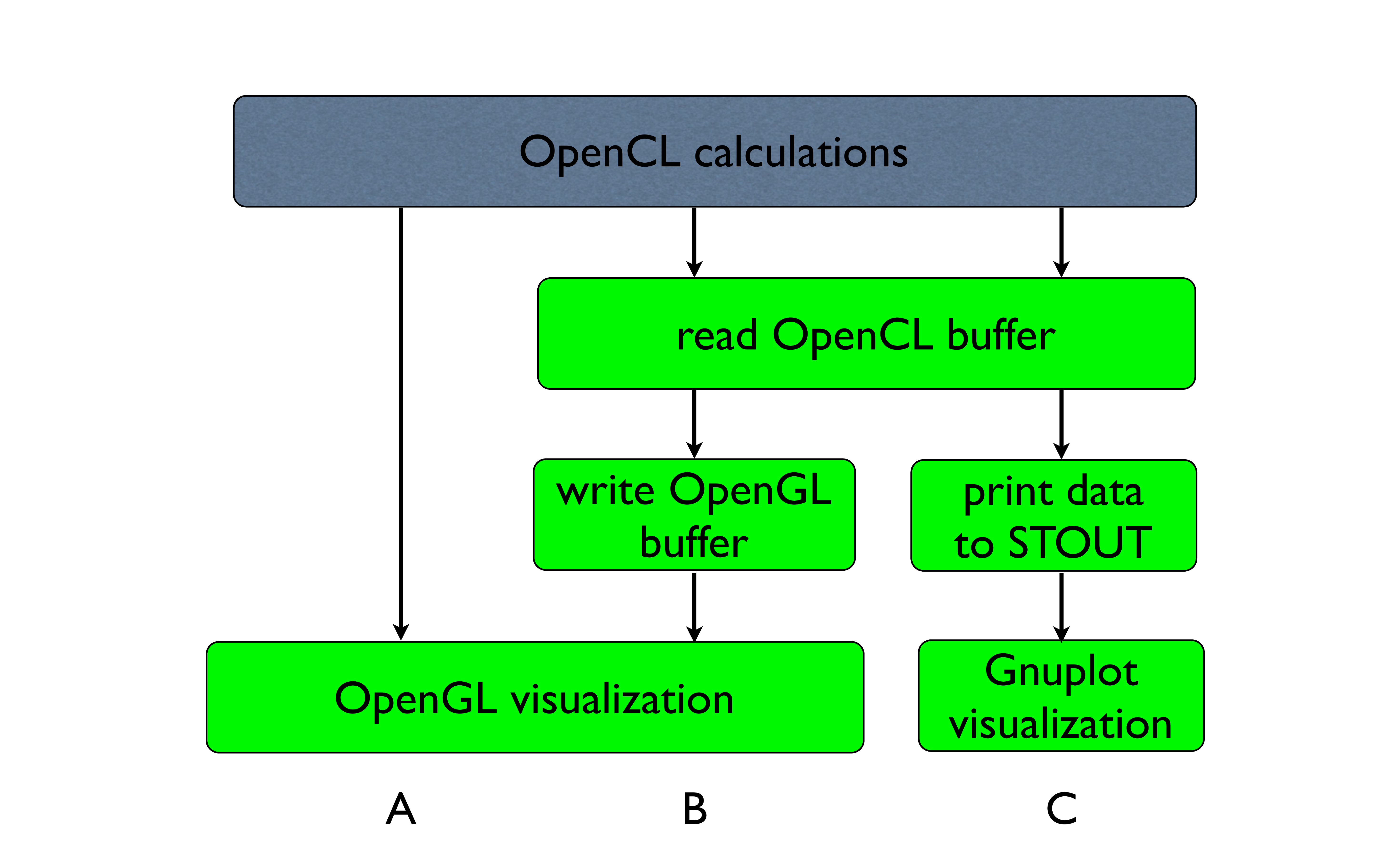}
\caption{Different approaches to OpenCL calculation visualization.
From left: OpenCL/OpenGL interoperation with shared buffers (a), OpenGL is used to visualize results, but buffers are not shared with OpenCL (b),  almost traditional approach where results are send to standard output, captured by gnuplot and visualize there (c).}
\label{schema_app}
\end{center}
\end{figure}

Three different approaches will be presented and compared (as pictured in the Fig. \ref{schema_app}).

\subsubsection{Presenting results with external software}

In the traditional approach calculations are carried out on a fast device (CPU, GPU), final results are saved into files and finally data are plotted with external software (e.g. gnuplot).
This procedure may be adequate if one carries long calculations with a single final plot.
The impact of memory transfer issues, time of an external plotting software to initialize, read data from a file and plot it is irrelevant comparing to the time of calculations.

\subsubsection{Intermediate steps with OpenGL}

In this strategy, both calculations and visualization of the results are done by the same program.
OpenCL device runs the kernel, in which calculations are done.
Results are saved into the host memory and then they are plotted onto the screen, using OpenGL.
OpenGL is initialized only once at the beginning of a program run.
 After receiving new data the screen is only updated.

\subsubsection{Intermediate steps with OpenGL/OpenCL shared buffers}

In the proposed method, there is no transfer of calculation results from OpenCL device to the host memory.
Both computation and visualization operate on the same buffers.

OpenCL and OpenGL specification requires that if one wants to use shared memory buffer in-between those two frameworks, first OpenGL buffer should be created.
Secondly, instead of creating plain OpenCL buffer ({\it clCreateBuffer}), it has to be created from the OpenGL one ({\it clCreateFromGLBuffer}).
In OpenGL world, buffers hold mostly either information about position or color.

For example, color buffer can hold a chain of float numbers representing colors in RGBA scheme.
The four numbers ($a_1 ...
a_4$) stand for three color intensities (red, green, and blue with $a_i \in [0,1]$) and $a_4$ stands for the alpha opacity controller.
Similarly, vertex buffer holds a chain of float numbers that every four represent position in homogeneous coordinates (x,y,z,w).

\subsubsection{Comparison of performance}

\begin{figure}[htbp]
\begin{center}
\includegraphics[scale=0.45]{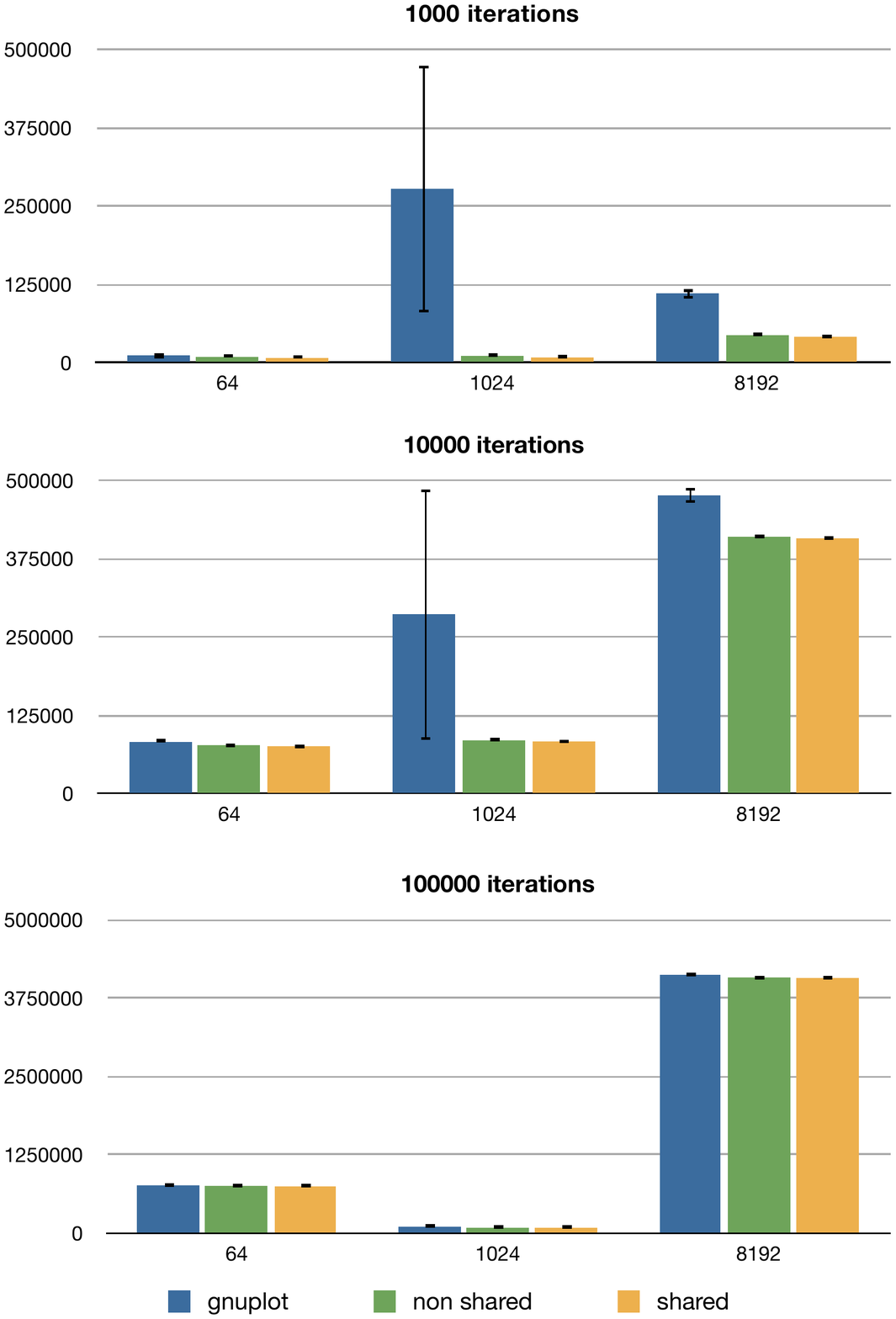}

\caption{Time in $\mu s$ of one calculation cycle  for the shared buffers method, non-shared buffers method and the reference gnuplot technique}

\label{fig_all_times}
\end{center}
\end{figure}

The time of one step has been calculated for every method for three different numbers of simulation iterations and for three different numbers of work itmes (starting parameter of the force $F_0$).
Resultant time has been averaged over 15 consecutive steps and plotted in the Fig. \ref{fig_all_times}.

For longer runs (with high number of iterations) the performance of all three methods seems to be comparable.
On the other hand, the lower number of iterations, the more striking is the  difference between different approaches.
It can be found that for given numbers of calculated parameters the time difference between methods is more or less constant.

\begin{figure}[htbp]
\begin{center}
\includegraphics[scale=0.3]{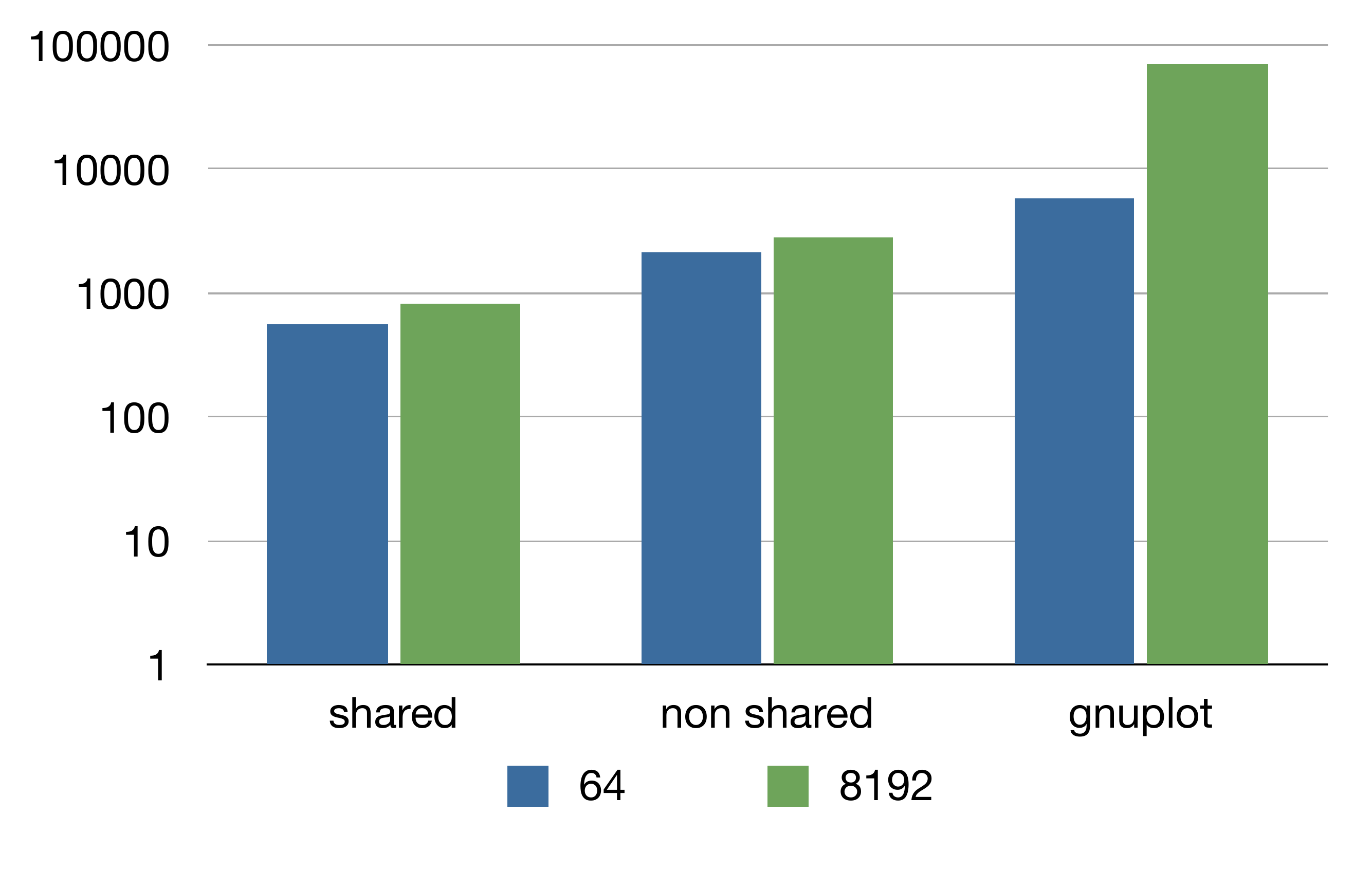}
\caption{Time in $\mu s$ consumed by every calculation step for data preparation, result acquisition and visualization.
Notice log scale for time.}
\label{fig_offset_comp}
\end{center}
\end{figure}

Using linear regression one can estimate the average time for one iteration for different methods and number of parameters with an offset being time of data preparation, result acquisition and visualization.
Results of the latter operation can be found in the Fig. \ref{fig_offset_comp}.

It appears that for gnuplot reference technique, time  dedicated to visualization can be longer by even two orders of magnitude comparing to the fastest shared buffers method.
On the other hand, non-shared buffers method can be a few times longer than the shared buffers approach.
For all tested methods of transferring data to the screen it has always taken more time to show plots of larger data.
However, that impact was much more evident for gnuplot approach than for other, OpenGL techniques.

One of the ways to present comparison between different visualization methods is to show the number of possible calculation steps that could have been done in time spent on a visualization step.
Data visible in the Fig. \ref{fig_offset_comp} has been divided by time of one iteration (that has been calculated from the same linear regression as mentioned before).
Results can be found in the Tab. \ref{tab_lost_time}.

\begin{table}[ht]
\tbl{Number of possible iterations that could be taken in time lost on visualization}
{\begin{tabular}{ccc}
\toprule
method & $2^6$ parameters & $2^{13}$ parameters \\
\colrule
share buffers & 80 steps & 20 steps\\
non-share buffers & 283 steps & 69 steps\\
gnuplot & 768 steps & 1734 steps \\
\botrule
\end{tabular}
\label{tab_lost_time}}
\end{table}%
 
For a longer calculation the number of possible iterations that could be taken in the time of visualization even for the slowest gnuplot reference method is negligible comparing to number of iterations done.
On the other hand, for relatively short runs (e.g. screening of data every $1000-2000$ steps) for larger sets of parameters visualization can take more time than calculation itself.

\section{Conclusions}

In this paper I have shown the approach to visualize OpenCL calculation of stochastic differential equations using co-existing OpenGL framework.
That technique simplifies the workflow of SDE calculations, without losing performance boost from graphic card use.

Presented OpenGL approach, especially with shared buffers, can help to gain better insight to calculation in real time.

Vendor agnostic OpenCL can be run on different devices, even rewriting naive C code and running on the same CPU can give outstanding improvement of a calculation time.

\bibliographystyle{iopart-num}
\bibliography{stepclbib}

\end{document}